\documentclass 
{aa}

\usepackage{aabib99}

\usepackage{graphicx}

\newcommand{\apjl}{ApJ Let.}
\newcommand{\apj}{ApJ}
\newcommand{\mnras}{MNRAS}

\newcommand{\aap}{A\&A}
\newcommand{\aaps}{A\&A Supp.}

\begin{document}

\title{Gamma Ray Burst progenitors -- a case for helium star mergers}

\titlerunning{GRB progenitors...}

\author{Krzysztof Belczy{\'n}ski \inst{1,2}, Tomasz Bulik \inst{1}, 
        Bronis{\l}aw Rudak \inst{3}}

\mail{K. Belczy{\'n}ski, kabel@camk.edu.pl} 
 
\authorrunning{Belczy{\'n}ski et~al.}
 
\institute{
 Nicolaus Copernicus Astronomical Center, 
 Bartycka 18, 00-716 Warszawa, Poland
\and
 Harvard-Smithsonian Center for Astrophysics,
 60 Garden St., Cambridge, MA 02138, USA
\and
 Nicolaus Copernicus Astronomical Center,
 Rabia{\'n}ska 8, 87-100 Toru{\'n}, Poland
}

\date{Received ..., Accepted ...}

\thesaurus{08.02.03, 08.05.03 -- 08.14.01 -- 13.07.01}
 
\maketitle 
 
\begin{abstract} 
{  

Recently much work in Gamma-Ray Burst (GRB) studies was devoted to    
revealing the nature of outburst mechanism and to looking for 
GRB progenitors.  Several types of progenitors were proposed for
GRBs.  Most promising objects are collapsars, compact object
binaries,  Helium star mergers and recently discussed 
supernovae. In this paper we consider four proposed binary star
progenitors  of GRBs: double neutron star (NS-NS), black hole
neutron star  (BH-NS), black hole white dwarf (BH-WD) mergers
and Helium  star mergers (He-BH). Helium star mergers are a
possible outcome of common envelope  evolution of a compact
object entering the envelope of a giant  with a helium core. 

Using population synthesis we calculate number of the binary
progenitors and show that BH-WD and Helium star mergers
dominate  population  of the proposed binary progenitors.
Comparison of distribution of different binary mergers around 
galaxies they are born in, with localization of GRB afterglows 
in their host galaxies shows that only Helium mergers may be 
responsible for GRBs with observed afterglows while it excludes 
NS-NS, BH-NS and BH-WD systems as GRB progenitors. Assuming that
all GRBs come from Helium star mergers and comparing  numbers of
Helium star mergers with observed BATSE GRB rate let us  derive
upper limit on GRB collimation to be $\sim 4^\circ$. } 
\end{abstract}


\section{Introduction}

The last decade  brought a great breakthrough in gamma-ray
burst  studies. Observations of GRB afterglows in X-ray, optical
and radio wavelength  domains 
\cite{1997IAUC.6576....1C,1997IAUC.6584....1G}  led to
identification of GRB host galaxies \cite{1997IAUC.6588....1G} 
and measurements of their redshifts. This has solved the long
standing problem of their origin. While we learned that GRBs
come from cosmological distances there  are still two major
difficulties in understanding this phenomenon. First, we do not
understand fully the physics of outburst. Although several models
were proposed they all yet have to  meet some severe
constraints imposed by observations (i.e. releasing  energies of
$10^{51}$--$10^{54}$\ ergs in timescales as short as 
$10^{-2}\,$s
in case of some GRBs). Second we don't know what are the
astrophysical objects leading to  gamma-ray bursts, i.e. what are
their progenitors?

Recently a black hole accretion disk model of GRB outburst has
been  favored \cite{Me99,Bl00,FWH99}. Progenitors leading to
this model include  collapsars \cite{Woosley93,MacFadyen99} and
binary stars:  Helium star mergers \cite{Fryer98},  double
neutron stars \cite{Ruffert97,Meszaros97},  black hole neutron
star \cite{Lee95,Kluzniak98} and  black hole white dwarf systems
\cite{Fryer99b}.  Also recently  the  supernovae  gamma-ray
bursts connection received much attention \cite{P99,W00,Ch00},
however there is  still no clear evidence that these two
phenomena are intrinsically  correlated
\cite{1999A&AS..138..469G}.

The present paper is an extension of our previous studies  
\cite{BB1998,BBZ99,BBZ2000} and we aim here at a discussion of
viability of the proposed GRB binary progenitors.  

One way of telling which group of proposed binaries might be 
responsible for GRBs is to predict their numbers and compare
them to the observed number of GRBs.  Population synthesis is a
powerful tool for predicting numbers of  binary populations
although it suffers from many uncertainties as some parameters
of single and binary evolution are poorly known.  Moreover,
population synthesis works well in predicting  the relative
numbers of events, while calculation of absolute numbers
requires additional assumptions. Using population synthesis
method we calculate production rates of Helium star mergers,
double neutron stars, black hole neutron star and black hole
white dwarf systems. We use BATSE detection limit of observed
number of GRBs to compare  with our predicted numbers of the
binary progenitors.    

Another way of discerning among the binary progenitors is to 
compare their merger site distributions around host galaxies
with localization of GRBs within host galaxies. Again this may
be accomplished by population synthesis method to  simulate a
given binary population and then placing it in  galactic
gravitational  potentials one may trace each binary until its
components merge due  to gravitational wave energy losses. We
perform such  calculations for different galactic potentials 
and then compare the results to the observations of GRB
afterglows and their  positions within host galaxies.  

In this work we extend our previous studies to include two more 
proposed binary progenitors: black hole white dwarf binaries
and  Helium star mergers. Moreover, we improve our population
synthesis code and for consistency  we also present the updated
results for two previously studied types of  proposed
progenitors: double neutron star and black hole neutron  star
systems.

Population synthesis method has already been applied to study
compact  object binaries in context of GRB progenitors. However
most of authors have concentrated only on double  neutron stars
and black hole neutron star systems
\cite{Nar91,1991ApJ...380L..17P,1993MNRAS.260..675T,1996A&A...312..670P,1999MNRAS.305..763B,BB1998,BBZ99,BBZ2000}.
Only one group \cite{FWH99} presented calculations including all
types  of proposed binary progenitors and also collapsars. Our
calculations besides the conclusions we obtain, may serve as a
direct check and  comparison for the  results presented by
Fryer et al., \cite*{FWH99}. 

In section 2 we present observational data on the distribution of
GRBs around their host galaxies, in section 3 we describe the
population synthesis code used in this paper and formation
scenario for Helium star mergers, in section  4 we discuss the
results, and we present our conclusions in section 5.

\section{Observations -- Distribution of GRBs}

The discovery of gamma-ray burst afterglows by the Beppo SAX
satellite  lead to identification of GRB host galaxies, and to
localization  of the GRB events in relation to these galaxies. 
A list of GRBs with afterglows and their projected distances
from the centers of host galaxies is shown in  Table~\ref{Loc}.

\begin{table}    
\caption[]{Localization of GRB afterglows in relation to the host galaxies}
\label{Loc}      

\begin{tabular}{l|l|l|l}
GRB     &  redshift $z$ &  Offset $\Delta \Theta$ & R$_\perp$\ [kpc]\\
\hline
970228  &  0.695        &  $ 0.30$''              & 3.8 \\
970508  &  0.835        &  $ 0.01$''              & 0.15 \\
971214  &  3.42         &  $ 0.06$''              & 2.1 \\
980703  &  0.966        &  $ 0.21$''              & 3.5 \\
980613  &  1.096        &  $ 0.8$''               & 15 \\
990123  &  1.61         &  $ 0.7$''(?)            & 16(?) $<5$ \\
990510  &  1.60         &  $ <0.08$''             & $<2$ \\
990712  &  0.434        &  $0.24$''               & 1.4 \\
\end{tabular}
\end{table}

From Table~\ref{Loc} we see that GRBs take place not far from
the  centers of their host galaxies and in the case of
GRB970508 the offset is only  $ 0.01$''  off the center of the
host galaxy. For a review of recent observations see
\cite{Bloom2000}.  Moreover, the host galaxies are
typically small,  irregular, with  intense star formation
processes \cite{Fru2000}. The data presented in Table~\ref{Loc}
describes only long GRBs as  only for these bursts afterglows
were so far observed.

\section{The model}

We use the population synthesis code described in detail in 
Belczy{\'n}ski \& Bulik \cite*{BB1998}. The code was modified to
allow for evolution of low mass and  intermediate mass stars.

\subsection{Modifications and description of population synthesis code}

{\bf Initial conditions.} All binary star initial parameters are
drawn from the same distributions,  but we have changed the
limits of binary components masses. As of now, we use more
accurate description of single stellar evolution (see below);
now we let the primary (more massive component at beginning) to 
have mass in range 8.6--100 M$_\odot$. The lower limit has been
set  in order to make sure that the primary will explode in a
supernova  explosion (unless it is stripped of mass in early
mass transfer/loss episode) and turn to a neutron star or a
black hole.  The range of secondary mass component is restricted
now to  1.0--100 M$_\odot$. The lower limit here ensures that
the star has a chance to produce a remnant  in the Hubble time.
This way we study population of binaries which have the most 
chances to produce the proposed binary star GRB progenitors.

\noindent {\bf Single stellar evolution.} To describe evolution
of single star we use analytical formulae  of Eggleton et al., \cite*{Egg89} and
Tout et al., \cite*{Tout96} which are a new addition to our code.  We follow a
given star through different stages of its evolution:  main
sequence, Hertzsprung gap, red giant branch, core helium
burning,  and asymptotic giant branch. The star may become a
naked Helium star due to wind mass loss and then  we follow its
evolution until it cools down to become a white dwarf or  if it
is  massive enough to explode as a supernova and become a
neutron star  or a black hole.

We introduce one important change to the formulae of
Tout et al., \cite*{Tout96}. In our calculations we assume that the mass of a
compact object formed in  a supernova explosion is half of the
helium core mass of exploding  star.  This results in the mass
range of compact objects stretching up to  $\approx 20 {\rm
M}_\odot$, which can be compared to the maximum mass of compact 
objects of $\approx 3 {\rm M}_\odot$ in the  original
prescription  of Tout et al., \cite*{Tout96}. We assume that compact objects
below 3 M$_\odot$\ are neutron  stars and over this limit black
holes (for discussion see Belczy{\'n}ski et al., \cite*{BBZ2000}).

Depending on its initial mass any star will turn to become
stellar  remnant, i.e. a white dwarf, a neutron star or a black
hole. During late stages of evolution (Hertzsprung gap through
asymptotic giant branch) stars are allowed to lose mass via
stellar wind. We follow Tout et al., \cite*{Tout96} and include standard
Reimers wind mass loss rate formula \cite{KR78} and for very
massive and large stars we include luminous blue variable strong
winds \cite{HD94}. It is important to note that Helium stars are
not allowed to loose  mass in our present code, and that leads
to overestimations of the  most massive compact object  masses.
Inclusions of Wolf-Rayet type winds for the naked Helium stars
would decrease our maximum compact mass down to $\approx$\
12--15 M$_\odot$. Such a change, however, would not change
qualitatively conclusions of the  present  study.

\noindent {\bf Binary evolution.} Binary evolution may change
entirely the evolution of any of its components. For components of
close binary systems,  which will interact during  the
course of evolution, we calculate mass loss/gain during mass
transfer/loss events. We include common envelope evolution,
quasi-dynamic mass transfer  and hyper accretion onto compact
objects during common envelope phases as in our previous studies
\cite{BB1998}. Components losing their Hydrogen-rich envelopes during
giant stages become  naked Helium stars, while those gaining mass
(most often during they main sequence life)  are rejuvenated.
Rejuvenation and naked Helium star evolution is treated as
described  by Tout et al., \cite*{Tout96}.

Any mass loss from the system (either through wind mass loss or during 
mass transfer/loss events) changes the binary orbit.
Following evolution of binary and its components we also include 
tidal circularization \cite{1996A&A...309..179P} and magnetic 
breaking \cite{Tout96}.

During supernova explosion we follow precisely the orbit
evolution  and we check if a system in which a supernova
explosion takes place survives  the event. We calculate the mass
of the newly formed compact object, the remaining mass is
expelled  from the system carrying off momentum, and a natal
kick is added to the newly formed compact object (either a
neutron star or a black hole) and  then the new orbit is
calculated. Systems which survive explosions receive additional
center of mass  velocity as an effect of the natal kick and mass
loss from the outbursting component. Once the evolution of binary
components is terminated and a system has survived mass
transfer/loss events and supernova explosions we study 
populations of proposed binary GRB progenitors, that is NS-NS,
BH-NS  and BH-WD compact object binaries. These binaries evolve
only due to gravitational wave energy loss  which will cause the
orbital separation decrease and finally lead to
a merging event of two compact components and possibly  to a
gamma-ray burst.    For supernova explosions and gravitational
wave energy loss we use  same treatment and formulae as
in Belczy{\'n}ski and Bulik, \cite*{BB1998}.

\subsection{Helium star mergers}

In this subsection we will describe more specifically an evolutionary 
path which may lead to formation of a Helium star merger, a new 
feature of our code.

The binary components may merge  during mass transfer  events,
provided that the  orbit is not too wide. As an example let us
consider a binary, with an small enough orbital separation, so
that  the primary (the initially more massive component) during
its expansion on the giant branch   overfills its Roche
lobe.  Let us suppose that the mass ratio (secondary to primary)
is small, so that the mass transfer  will proceed on the 
dynamical timescale. The entire envelope of the giant primary
will be lost, and it will  become a naked Helium star.  The
secondary which is still on its main sequence will not have the
time to accept  any of the matter shed from the primary
envelope, so it will survive the mass loss  virtually unchanged.
As a consequence of this mass loss event the binary orbit will
shrink  drastically. If there is enough orbital energy to expel
the entire envelope of the primary  then the system will survive
(otherwise the helium core will merge with main sequence star, a
case we are not interested in this study) and continue  its
evolution. If the Helium star is massive enough it will undergo
a supernova Ib type explosion.  Since the orbit is tight after
the first mass transfer  there is a good chance that  the system
will survive the supernova explosion.   Thus the  system now
consists of a neutron star or a black hole and a main sequence 
star.  As a consequence of mass loss from the system and
randomly added kick to newly formed compact object the orbital
separation  increases and the orbit  becomes eccentric.   As the
time goes on, the secondary, which is probably a more massive 
component now, will start its evolution up the giant branch.
Once the secondary radius approaches the Roche lobe,  tidal
interaction  circularizes the orbit and decreases the orbital
separation.  When the secondary overfills its Roche lobe the
system  goes  through a similar phase as during the first mass
transfer.  If, as we have mentioned above, the secondary  is the
more massive component then the  mass transfer proceeds again on
dynamical time  scale.  In this mass transfer phase  when the
envelope of the secondary engulfs the system in a common
envelope the compact object begins to spiral in toward  the helium
core of the giant.  There is a chance now that the compact
object will accrete some material from the envelope,  as pointed
by Bethe and Brown, \cite*{Bethe1998}, and if it is a  neutron star it may
collapse to form a black hole.  Since the orbit is already tight
when the system enters the second mass transfer phase the
orbital energy available is not sufficient to expel the envelope
of the secondary.  The spiral in continues until the black hole
enters Helium core of the giant. It disrupts tidally the Helium
core, swallowing at the beginning a part of it,  but the
remaining  helium material will form a hot, rapidly rotating
disk around the black hole \cite{Fryer98}.  This configuration,
a Helium star merger, may lead to a gamma-ray burst  in the  black
hole accretion disk GRB model \cite{FWH99}.

\section{Results and discussion}

\subsection{Relative production rates}

\begin{figure}
\includegraphics[width=0.45\textwidth]{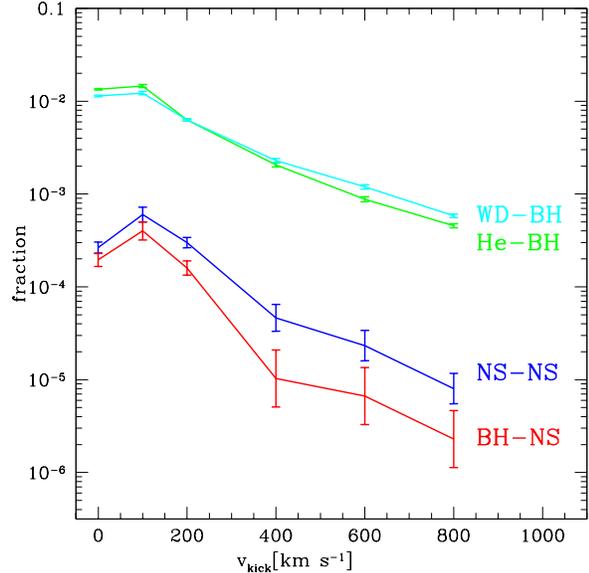}
\caption{
Relative fraction of different GRB progenitor types that
merge (or manage to evolve) within the Hubble time.}
\label{frac}
\end{figure}

In Figure~\ref{frac} we show the relative numbers of four
different GRB  progenitor types that merge within the Hubble
time (15 Gyrs) as a  function of the width of the distribution
from which we draw kick  velocity a compact object receives in a
supernova explosion.

Two things are clearly seen; first the number of WD-BH binaries 
and Helium mergers (He-BH) is about the same and is more then
an  order of magnitude greater then the number of NS-NS and
BH-NS  binaries.  It means that if GRBs originate from binary
mergers then they  mainly come from WD-BH binaries and/or Helium
mergers.

Second, the relative number of a given progenitor type falls
off  approximately exponentially with the kick velocity. This is
quite clear, the larger kick compact object receives the  larger
chance that the system will be disrupted in supernova 
explosion. And this is a reason for smaller production rates
with  increasing kick velocity. Assumptions about the kick velocity
distribution plays an important role  on results of compact
object binaries population synthesis. We draw the kick
velocities from a three dimensional Gaussian  distribution of a
given width which we treat as a parameter in  our studies. There
is a line of evidence coming form pulsars galactic velocity 
observations that the distribution of kicks is bimodal and 
consists of a weighted sum of two distributions: about 80\%
are   intermediate kicks of about 200 km s$^{-1}$ and the other
20\% are high velocity kicks of about 700 km s$^{-1}$ 
\cite{1997ApJ...482..971C}. However if real distribution is in
fact bimodal, then   the high velocity component won't have
significance for  properties of the population of  compact
object binaries, as these will tend to be  disrupted by high
velocity kicks.  High velocity component will imprint its
presence in the kick  distribution through high peculiar
velocities of single pulsars and  will decrease number of
compact object binaries.    The lesson from this is that if one
studies properties of compact  object binaries, and not their
numbers, then only the lower component  of the kick velocity
distribution is relevant.

Relative production rates may be calibrated to obtain the
real rates in our Galaxy(eg. see eq. 14 in 
Belczy{\'n}ski and Bulik, \cite*{BB1998}). For example, for the width of kick velocity of 
v$_{kick}=200$ km s$^{-1}$ we obtain: 1 merging event per Milky
Way like galaxy per 10$^6$\ yrs for BH-NS  systems, 3 events for
NS-NS binaries and 60 for WD-BH and Helium  mergers. These
numbers are obtained under assumption that binary fraction is 
50\%, that there are 0.02 supernovae per year in the Galaxy, and
that  the star forming process has been constant throughout the
history  in the Milky Way.

\subsection{History of binary merging events}

\begin{figure}
\includegraphics[width=0.9\columnwidth]{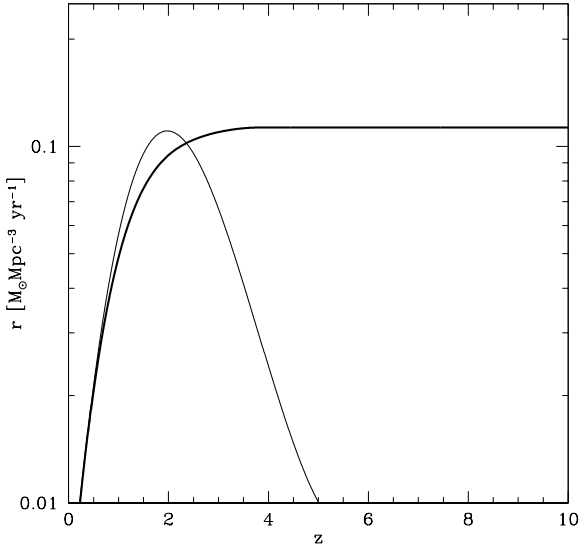}

\caption{Star formation history rates used in this work. The thin
line is based on Madau et al., (1996), 
while the thick line represents
approximately the rate of Rowan-Robinson (1999).}
\label{sfr}
\end{figure}

The results of the population synthesis code can be combined
with  our knowledge of the star formation rate history to yield
the  rate of various types of GRB progenitors as a function of
redshift. Star formation history at high redshift is not well
known, however it is generally agreed that the star formation 
rate rises steeply up to  $z\approx 1$. At higher redshifts  the
analysis of the Hubble Deep Field   \cite{Madau96} provided 
lower limits on the rate, yet these limits decrease with
increasing redshift.On the other hand Rowan-Robinson  
\cite*{1999Ap&SS.266..291R} argues that the star formation 
does not fall down and remains roughly at the same level above
$z=1$. We consider two cases: a star formation function falling
down steeply above $z\approx 1$ (the thin line in
Figure~\ref{sfr}), and a case of strong star formation continuing
up to $z=10$ (the thick line in Figure~\ref{sfr}).

For a given type $i$ of the GRB progenitor we can calculate 
the number of events up to the redshift $z$:
\begin{equation}
N_i(>z) = 4\pi \int_0^z r_z^2 {dr_z\over dz} {R_i(z)\over 1+z}
dz\, ,
\end{equation}
where $r_z$ is angular size distance $r_z = c H_0\int_0^z
(\Omega_m(1+z^3)+\Omega_\Lambda)^{1/2}$. $R_i(z)$ is the rate of a
given type of event at the redshift of $z$:
\begin{equation}
R_i(z) = \int_{t(z)}^{t'}  R_{sfr}(z)*f_i*(t')p(t-t') dt' \, ,
\end{equation}
where $t$ is the conformal time, $dt =dz (1+z)^{-1}
((1+\Omega_m)(1+z)^2 - z(z+2)\Omega_\Lambda)^{-1/2}$,
$p(t)$ is the probability distribution of a merger of a given type
as a function of time since formation of the system.
We calculate the distribution $p(t)$ for each type of a merger
using the  population synthesis method.

\begin{figure*}
\includegraphics[width=0.48\textwidth]{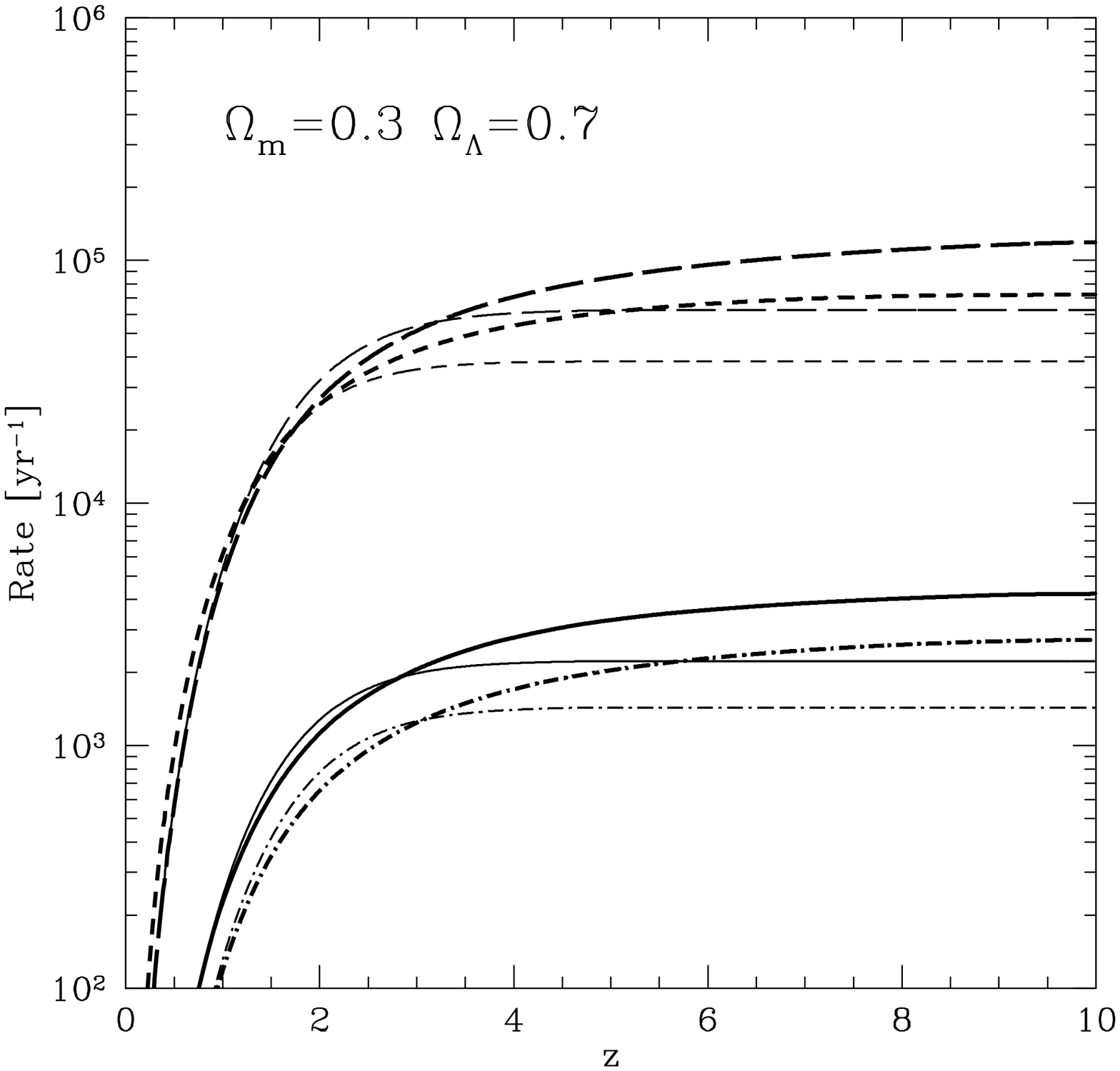}
\includegraphics[width=0.48\textwidth]{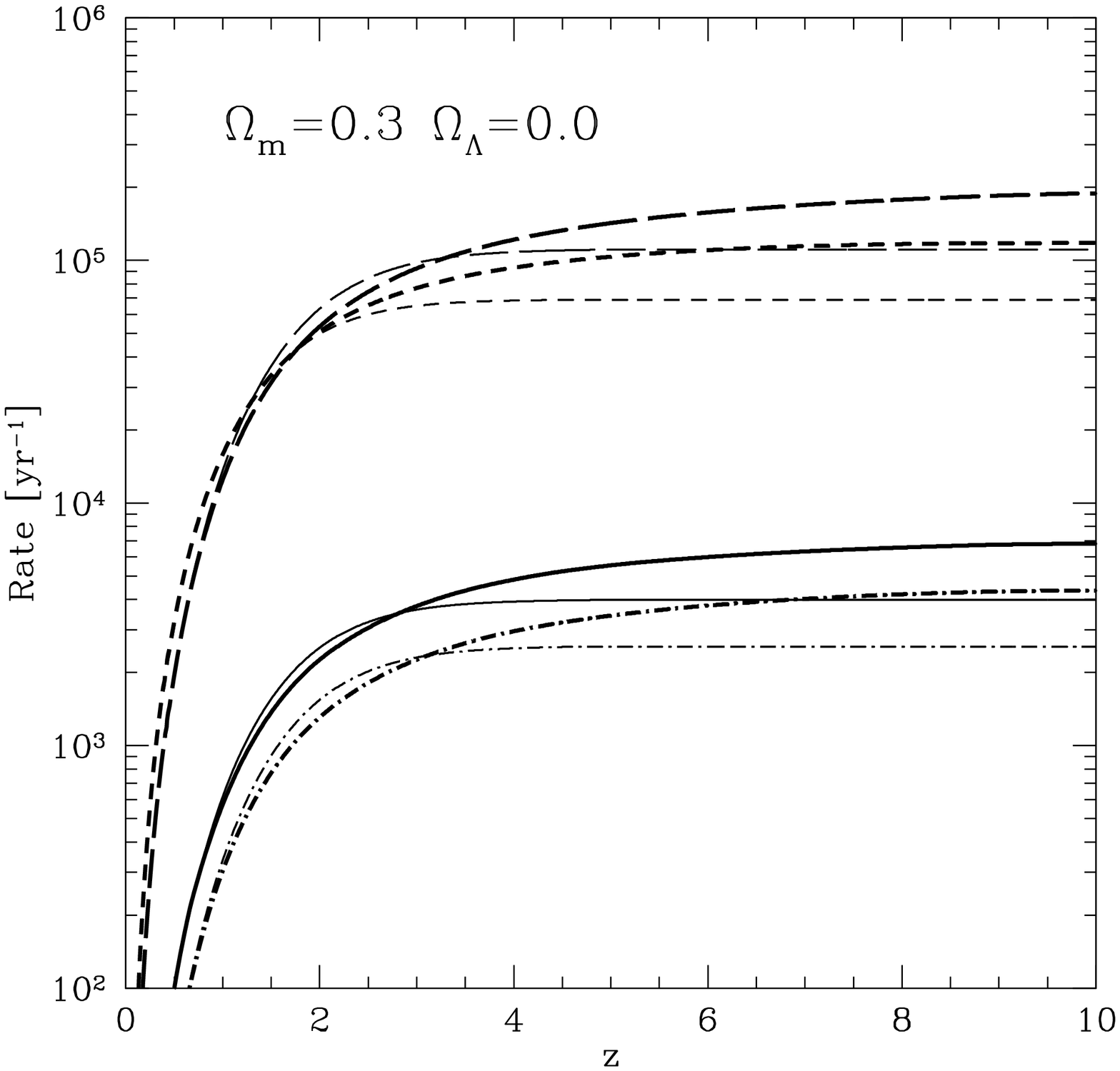}
\caption{
Cumulative event  rates of different types of GRB
progenitors. The left panel corresponds to the cosmological model
described by $\Omega_m=0.3$ and $\Omega_\Lambda=0.7$, while the
right panel to $\Omega_m=0.3$ and $\Omega_\Lambda=0$. 
The solid  line describes the NS-NS rate, the dash dotted line
describes the BH-NS rate, the short dashed line corresponds to
the BH-WD rate and the long dashed line shows the He-BH rate.
The thin lines corresponds to the assumed star formation rate of
Madau et al., (1996), 
while the thick lines represents the results
obtained using the  rate of Rowan-Robinson (1999).
Note that BH-WD, and He-BH events strongly dominate the population.}
\label{zdist}
\end{figure*}

In Figure~\ref{zdist} we show the cumulative rates of different 
merging events as a function of redshift. We have combined our
relative numbers (shown on Figure~\ref{frac}) for different
progenitors with the star formation rate function 
\cite{Madau96} and \cite{1999Ap&SS.266..291R}, 
and after taking into account the 
evolutionary time delay of a given merging event we integrated
our relative production rates to find the merger rates as a
function of  redshift. In this example calculation we used
two cosmological models with and without the cosmological constant:  
  $\Omega_M=0.3, \,\,
\Omega_\Lambda = 0.0 $, and  $\Omega_M=0.3, \,\,
\Omega_\Lambda = 0.7 $.
In both cases  the Hubble constant is $H_0 =65$ km
s$^{-1} {\rm Mpc}^{-1}$. We used the kicks drawn from the
distribution which is a weighted  sum of two Gaussians: 80
percent with the width of 200 km s$^{-1}$ and 20 percent with
the width 800 km s$^{-1}$.

The curves in Figure~\ref{zdist} can  be  compared with the 
BATSE gamma-ray burst detection rate  corrected for BATSE sky
exposure, which is $\approx\ 800$ events per year. Comparison of
the cumulative distributions for different progenitor types with
the BATSE rate shows that if any of the progenitor types
included  in our calculations were to reproduce the BATSE rate,
then we should not  see GRBs from redshifts greater then
$z=0.3$! Of course this is not the case (e.g. see
Table~~\ref{Loc}), as GRBs with  higher redshifts were observed,
and the median observed  GRB redshift is  $z \approx 1$.
However, we have not yet introduced the collimation factor into
our  results. The predicted cumulative rates presented in
Figure~\ref{zdist} will  decrease if we account for collimation
and thus restricted visibility of gamma-ray bursts. To lower
down our calculated rates to the BATSE rate, for average GRB
redshift of about unity, we would need the collimation of about
$4^\circ$ for BH-WD and Helium mergers and about $12^\circ$ for
BH-NS and NS-NS mergers. And, as the population is dominated by
BH-WD and Helium mergers then overall requirement for
collimation is to be $\approx 4^\circ$. This is only an upper
limit as we would expect such a collimation factor if  all GRBs
were originating only from binary progenitors.

The thin lines in Figure~\ref{zdist}  flatten out for high 
redshifts (z $\geq$ 5). In other words we do not expect binary
mergers at high redshifts. This is a combined effect of two
factors. First,  the star formation rate function (SFR) we have
used  \cite{Madau96,Totani97} falls down steeply for high
redshifts. This means that at high redshifts we do not expect
many stars, and thus their mergers. However the SFR we have used
is highly uncertain for high redshifts  and our result here may
be quantitatively questionable. However, the thick lines in
Figure~\ref{zdist}, corresponding to the predictions of the 
SFR of Rowan-Robinson \cite*{1999Ap&SS.266..291R} do climb up with redshift.
In this case the number of GRBs up to the redshift of  $10$ is
nearly double that up to the redshift of $2$. Thus, future 
detection (or non detection) of GRBs from such high redshifts 
will serve as a probe of the SFR at large redshifts.

Non zero lifetimes of binary progenitors are the second thing that 
makes our curves to flatten out with redshift.
Binary GRB progenitors need a specific time to evolve to a
compact object binary or to a Helium merger (t$_{evol}$) and the
compact object binaries need  time to merge due to gravitational 
wave energy losses (t$_{merger}$).
This times are non negligible and are specific for each group of 
proposed binary GRBs progenitors. 
For our sample of binaries we found characteristic lifetimes which 
are the sum of the evolutionary times and merging times 
$t_{life}$\ ($t_{evol}+t_{merger}$).
They are, 
for NS-NS: $\sim 10^7$--$10^{12}$\ yrs,
for BH-NS: $\sim 10^7$--$10^{10}$\ yrs,
for He-BH: $\sim 10^6$--$10^9$\ yrs,
for BH-WD: $\sim 10^7$--$10^{12}$\ yrs.
We see that these times are non negligible, and even if star 
formation process has begun at some point, for a given $z$, 
we need to wait at least 10$^6$--10$^7$\,yrs to start producing 
  GRBs of binary origin. 
 
An additional  point to emphasize here is that the
evolutionary times (t$_{evol}$), besides merging times, may 
play important role for some types of binary GRB progenitors.
They are less important for NS-NS and BH-NS systems which are 
end products of high mass stars (very fast evolution) and for 
which merging times (t$_{merger}$) are comparable with their 
total lifetimes $t_{life}$. However evolutionary times are
important for BH-WD systems, as it  may take a long evolutionary
time to form a white dwarf, usually  comparable with the merging
time of final compact black hole white  dwarf binary. It is even
more clearly seen in the case of Helium mergers for which  
evolutionary times equal total lifetimes, as these systems
merge  during common envelope evolution of still unevolved  (not
a compact object) binary. 

The above example calculation shows that the WD-BH and He-BH
events  are far more numerous than the NS-NS or BH-NS events. 
The absolute numbers presented above depend on a number of
assumptions leading from the binary population synthesis to the
observed rate.   The second rather robust result is that
if the star formation rate remains high even at high redshifts, 
the number of GRBs from such high $z$ must be significant - see
the difference between thin and thick lines in Figure~{zdist}.

\subsection{Distribution of binary mergers around host galaxies}

\begin{figure*}
\includegraphics[width=0.95\textwidth]{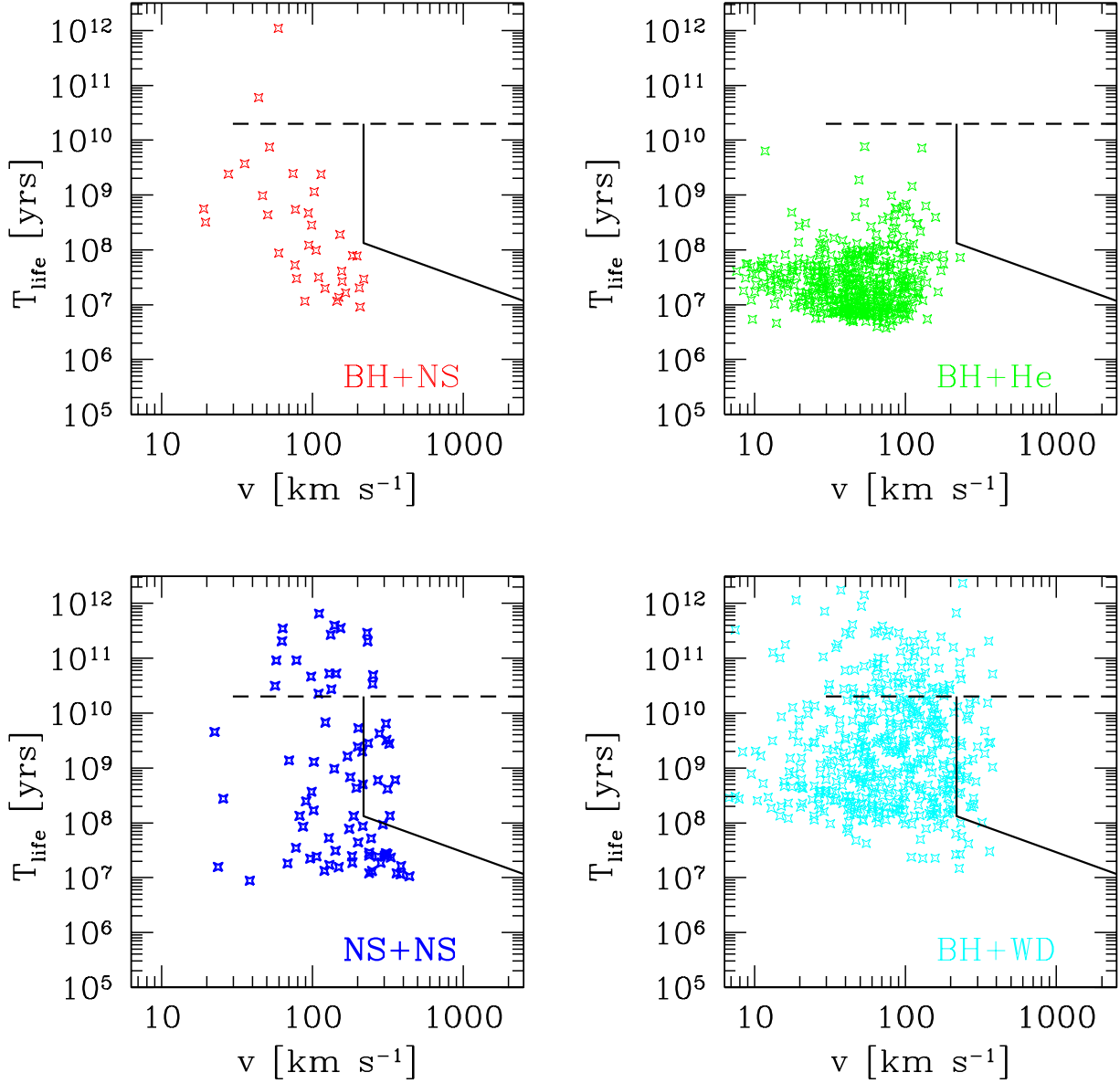}
\caption{ 
The distribution of different type of compact object binary
mergers in the space spanned by the binary center of mass velocity and
the binary lifetime (from ZAMS to final merger).
The horizontal dashed line corresponds to the Hubble time (15 Gyrs).
In the region for $t_{merge} < 15\,$ Gyrs we present two solid lines:
the vertical corresponding to $v=200\,$km~s$^{-1}$ -- approximately the
escape velocity from a galaxy, and the line corresponding to a constant
value of $v\times t_{merge} =30\,$kpc. Together these lines define the
region in the parameter space with systems that can escape from the host
large galaxy.}
\label{dots}
\end{figure*}

In Figure~\ref{dots} we present the distribution of center of
mass  velocities gained by systems in the supernova explosions
versus  binary lifetimes (the time binary takes to evolve from
ZAMS to final  merger of two components). Data is presented on
four panels for four types of the binary GRB  progenitors. On
each panel we plot three lines to show the distribution of 
mergers in respect to host galaxy of the size and mass
comparable  to Milky Way.  Together these lines define the
region in the parameter space  with systems that can escape from
a  massive galaxy \cite{1999MNRAS.305..763B}.

The horizontal dashed line corresponds to the Hubble time (15
Gyrs); we are not interested in the system above this line as
they do not  have a chance to merge within Hubble time.  

The vertical solid line corresponds to $v=200\,$km~s$^{-1}$,
which  is approximately the escape velocity from an intermediate
mass galaxy. Systems to the right of this line (with velocities
higher then  the escape velocity) will merge outside their
host galaxy.  

The inclined  solid line corresponds to a constant value of 
$v\times t_{merge} =30\,$kpc, which is about the radius of a  
high mass galaxy. Systems above this line have gained high
enough velocity  and have enough time to escape from their host
galaxy to  merge outside of it. 

We note that a significant fraction of NS-NS and BH-WD merging
events  takes place outside of  the host galaxies, and thus their
distributions are  inconsistent with the GRB observations.  This
conclusion is even stronger because it was drawn for the case 
of massive large host galaxy. As noted in Sec.~2, GRBs' host
galaxies are small, and in this case  even more NS-NS and BH-WD
mergers would take place outside their hosts then  it is
inferred form Figure~\ref{dots}.  One may argue that binary
mergers taking place outside galaxies in  a thin intergalactic
medium would not produce afterglows, and thus  would not
contribute to observed distribution of GRBs in relation to 
their host galaxies. However according to Costa \cite*{Costa00} all
GRBs observed by Beppo SAX  are accompanied by afterglows.

\begin{figure}
\includegraphics[width=0.45\textwidth]{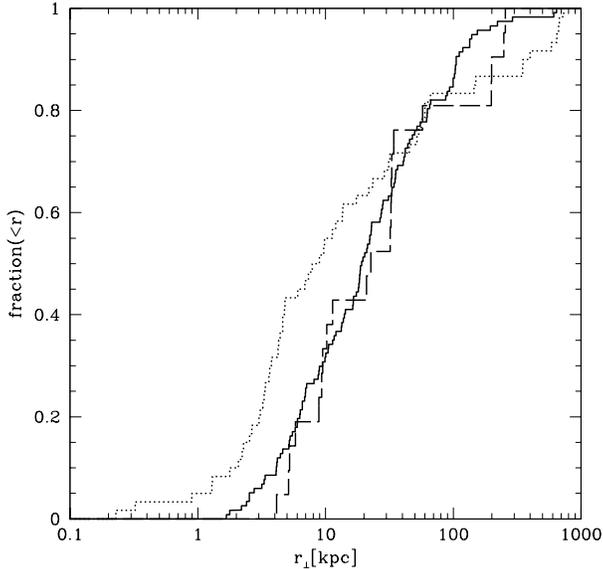}
\caption{
Cumulative distribution BH-NS events around low mass host 
galaxies. The solid line corresponds to the case of no
natal kicks $\sigma_v = 0$km\,s$^{-1}$,
the dotted line shows the case of $\sigma_v = 100$km\,s$^{-1}$,
and the dashed line represents the case of  $\sigma_v =
200$km\,s$^{-1}$.
More than 50\% of BH-NS merger events take place outside the 15 
kpc radius from the host.}
\label{kuma}
\end{figure}

\begin{figure} 
\includegraphics[width=0.45\textwidth]{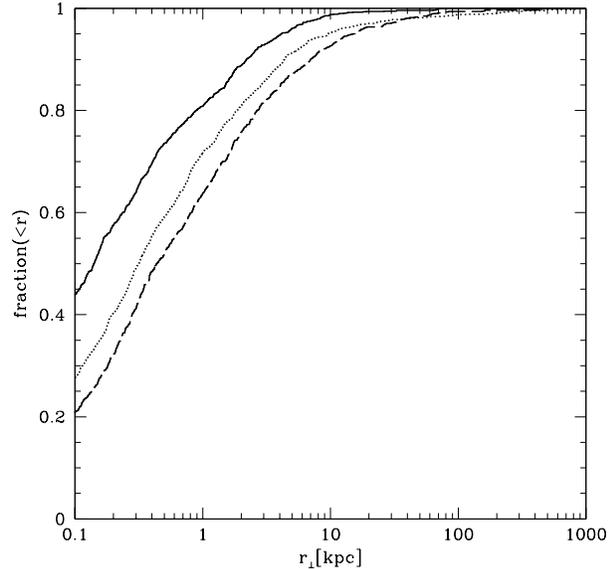}
\caption{
Cumulative distributions of Helium star mergers (He-BH) events 
around low mass host galaxies. The solid line corresponds to the case of no
natal kicks $\sigma_v = 0$km\,s$^{-1}$,
the dotted line shows the case of $\sigma_v = 100$km\,s$^{-1}$,
and the dashed line represents the case of  $\sigma_v =
200$km\,s$^{-1}$.
Almost 95\% of He-BH events takes place within the 15 kpc 
radius from the host.}
\label{kumb}
\end{figure}

Thus, we are left with two binary progenitors: BH-NS systems
and  Helium star mergers. As seen from Figure~\ref{dots} we
predict that all of them would  produce gamma-ray bursts inside
their host galaxies provided that  hosts are large and massive.
But as noted before this is probably not the case, and we have 
calculated their distribution in case when their hosts were
small, low mass galaxies. We approximate trajectories of our
systems in a potential of a small galaxy by  propagation in
empty space.  We place them in one point and let them move with
velocities  gained during supernova explosions. We follow their
trajectories until they finish their life in a  merging event.
We present the expected cumulative distributions of the
projected distance from the center of the host galaxy for BH-NS
merging events  in Figure~\ref{kuma}  and  for Helium star
mergers in Figure~\ref{kumb}.

In Figure~\ref{kuma} we see that more than 50\% of BH-NS
systems  merge outside the 15 kpc projected radius. This is in
clear disagreement with the GRB distribution  observations,
which show that all so far observed GRBs take place  within 15
kpc radius off their host galaxy center.     This excludes BH-NS
systems as potential GRB progenitors given  that GRB host
galaxies are small  \cite{Fru2000}.

Thus the only possibility  left for the binary origin of GRBs
are  Helium star mergers.  From Figure~\ref{kuma} we see that
almost 95\% of these merging events  take place within the
$15\,$kpc radius from the host, just as  expected if they
were gamma-ray burst progenitors! This result is independent of
the host galaxy size and mass, as we  have shown that these mergers
will merge within 15 Kpc radius even in the  case of no pull
from their host -- in the case of propagation in  empty space.

Now, if we  assume that all GRBs are coming from binary 
stars  we must remark that all long GRBs result from Helium star 
mergers and we estimate collimation factor of these burst to be 
$\sim 4^\circ)$. All the other merging events, coming form
NS-NS, BH-NS and BH-WD  systems may be responsible for short
GRBs, for which afterglows and thus  their distributions in respect
to host galaxies have not yet been  observed.

\section{Conclusions}

We have calculated the properties of the possible binary GRB
progenitors: BH-NS, NS-NS, BH-WD, and He mergers.  The  GRB
binary progenitor production rates fall off exponentially with
width of natal kick velocity distribution. We calculate the
expected redshift distributions and numbers for each type of the
progenitor, and find that BH-WD and He-BH type events dominate
over the NS-NS or BH-NS mergers. Moreover,  in calculating the 
redshift distribution of binary progenitors of GRBs one can not
neglect the evolutionary times $t_{evol}$. In our example
calculation we find that  assuming that all GRBs result from
binary mergers, then the population is dominated by BH-WD and
He-BH star mergers, and the collimation must be of order {$\sim
3 \times 10^{-3}\ (\sim 4^\circ)$}. The existence or non
existence of GRBs at high redshifts depends on the 
star formation history. In particular the recent measurement of
$z=4.50$ in case of GRB000131 \cite{Andersen2000} suggests
that the star formation rate has been high up to high 
$z$.

We find that only the  He-BH (Helium star merger) model of GRBs
is consistent with Beppo SAX observations of long bursts and
their localization in host galaxies. As noted in several previous
papers a larger fraction of NS-NS binaries merge outside even
massive host galaxies. The distribution of BH-NS mergers is
tighter around the massive  galaxies, however we find that they
do escape from the potentials of small galaxies. Our population
synthesis code results indicate that  a significant fraction of
the BH-WD binaries should obtain  large enough velocities so
that they can escape from the potential well even of a massive
galaxy. This differs  from the results of Fryer et al., \cite*{FWH99}, who
find
that the  BH-WD binaries should lie within the host galaxies. 
Moreover our code results in a much larger fraction of  BH-WD
binaries than the fraction Fryer et al., \cite*{FWH99} obtain. Finally we
remark that  NS-NS, BH-NS and BH-WD mergers can be responsible 
for short bursts which are not observed by Beppo SAX.  This
could be resolved by  HETE-II, if short burst afterglows are
discovered, and if their host galaxies are identified.

\acknowledgements We would like to thank Chris Fryer for helpful
discussions and comments during the 5th Huntsville GRB
Symposium. This work has been supported by the KBN grants
2P03D02219 (KB), 2P03D00418 (TB), and 2P03D02117 (BR)  and also
made use of the NASA Astrophysics Data System.

\end{document}